# Measurement of the neutron lifetime with ultra-cold neutrons stored in a magneto-gravitational trap

V.F. Ezhov,[1,5] A.Z. Andreev,[1] G. Ban,[2] B.A. Bazarov,[1] P. Geltenbort,[3] A.G. Glushkov,[1]
V.A. Knyazkov,[1] N.A. Kovrizhnykh,[4] G.B. Krygin,[1] O. Naviliat-Cuncic,[2,*] and V.L. Ryabov[1]

[1]*Petersburg Nuclear Physics Institute NRS KI, Gatchina, Russia*
[2]*LPC Caen, ENSICAEN, Université de Caen Basse-Normandie, CNRS/IN2P3, Caen, France*
[3]*Institut Laue-Langevin, Grenoble, France*
[4]*Institute of Electro-Physical Apparatuses, Gatchina, Russia*
[5]*Saint-Petersburg State University, Saint-Petersburg, Russia*
(Dated: December 23, 2014)

We report a new measurement of the neutron lifetime using ultra-cold neutrons stored in a magneto-gravitational trap made of permanent magnets. Neutrons surviving in the trap after fixed storage times have been counted and the trap losses have continuously been monitored during storage by detecting neutrons leaking from the trap. The value of the neutron lifetime resulting from this measurement is $\tau_n$ = (878.3±1.9) s. It is the most precise measurement of the neutron lifetime obtained with magnetically stored neutrons.

Precision measurements of the neutron lifetime provide stringent tests of the standard electroweak model [1] as well as crucial inputs for Big-Bang nucleosynthesis (BBN) calculations [2]. When combined with measurements of other neutron beta decay correlation coefficients [1], the neutron lifetime enables the determination of the $V_{ud}$ element of the Cabbibo-Kobayashi-Maskawa quark mixing matrix, providing a complementary unitarity test to that obtained from superallowed nuclear beta decay [3]. The neutron lifetime is also one of the key parameters for the determination of yields of light elements in BBN since the ratio between the free neutron and proton abundances drives the extent of fusion reactions during the first few minutes of the Universe [2].

The present world average value of the neutron lifetime as quoted by the Particle Data Group (PDG), $\tau_n$ = (880.3±1.1) s [4], is dominated by results obtained using ultra-cold neutrons (UCN) in material bottles. These results, and in particular the most precise of them [5], appear to be systematically lower than results obtained using a neutron beam and counting trapped protons following neutron decay [6]. A detailed discussion of the experimental methods and results can be found in Ref. [7].

The large discrepancy between the results indicates that all systematic effects are not fully under control. The importance of the neutron lifetime in particle physics and cosmology calls for alternative measuring techniques, with high sensitivity but other potential sources of systematic effects. We report here a new measurement of the neutron lifetime using UCN stored in a magneto-gravitational trap made of permanent magnets.

The repulsive force resulting from the interaction between the neutron magnetic moment and a magnetic field gradient can be used for the confinement of neutrons provided their energies are sufficiently low [8]. This has been incorporated for the measurement of the neutron lifetime in various configurations, the most successful having been a sextupole storage ring [9], leading to $\tau_n$ = (877±10) s, an Ioffe-Pritchard three dimensional trap leading to a storage time $\tau_S$ = (833$^{+74}_{-63}$) s [10], and an asymmetric Halbach array trap, with a storage time $\tau_S$ = (860±19) s [11].

The experimental setup used in the present measurement (Fig. 1) was operated at one of the beam positions of the UCN source PF2 at the Institut Laue-Langevin (ILL) in Grenoble. It comprises five main parts: a lift to fill the trap; the magnetic trap; a solenoidal magnet with a yoke; an outer coil around the magnetic trap; and the UCN detector. The central element of the setup is the magneto-gravitational trap made of NdFeB permanent magnets sandwiched between FeCo poles to generate twenty-poles. The trap is a vertical cylinder open at the top with a conical lower part open at the bottom. The central magnetic field generated by the poles is horizontal and the field gradient near the magnet surfaces is about 2 T/cm when moving toward the vertical axis of the trap. The trap is wrapped with an external coil to eliminate zero field regions in the trap volume. The magnets surfaces were covered with Fomblin grease (UT18 type) in order to reflect those neutrons which are not repelled by the magnetic field gradient and hit the magnet surfaces. Other technical details about the trap properties and design have been reported elsewhere [12,13].

A crucial aspect for the storage of UCN in magnetic traps is the filling of the trap. In previous experiments with this trap, the filling was performed from the bottom [12,13] through the magnetic shutter. This method suffers from serious shortcomings since neutrons are accelerated by the magnetic field gradient produced by the shutter. For the measurements presented here, a cylindrical lift located above the trap was used (Fig. 1). The cylinder is made of aluminum and its inner surfaces were covered with Fomblin grease. A disk of polyethylene was mounted inside the cylinder at an adjustable height to absorb UCN with energies above a given cut-off. The bottom cup of the lift can be separated from the cylindrical part for filling and emptying the lift volume. Below the trap, the solenoid is used as a fast magnetic shutter to close and open the trap. The counting of UCN is performed with a ³He detector, located 47 cm below the magnetic shutter, having a 100 μm thick aluminum entrance window. The lift and the trap are contained inside a vacuum chamber where the typical pressure was 1.2×10⁻⁴ Pa.

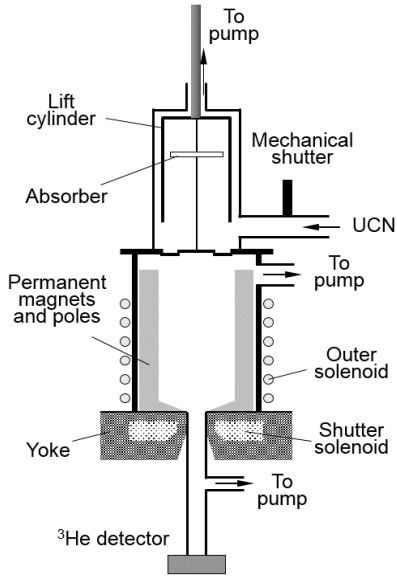

FIG. 1: Vertical section of the experimental setup showing the main parts: the lift (while being filled), the magneto-gravitational trap, the solenoidal magnetic shutter with its yoke, the outer solenoid coil and the UCN detector.

At any stage during a cycle, neutrons escaping the trap can be monitored with the ³He detector. Since the UCN are initially unpolarized, half of them cannot be stored after filling and escape through the bottom towards the detector. This provides a unique possibility to measure the number of UCN at the beginning of each filling. During the experiment the neutrons detected during the first 120 s were used for monitoring and data normalization. Other details about the trap filling sequence, including the conditions required to avoid heating of the neutrons, have been reported elsewhere [14]. Once the trap was filled, neutrons were stored for fixed storage times, after which the magnetic shutter was switched off to count the neutrons remaining in the trap. The neutrons were counted during "emptying" intervals of fixed duration. The background was then measured before starting a new cycle. In the measurements reported here, the counting of neutrons was integrated over 1 s time bins.

There are two possible sources of UCN losses in magneto-gravitational traps namely, the losses due to the flip of the neutron magnetic moment relative to the direction of the magnetic field and the losses due to the up-scattering of UCNs by the residual gas in the chamber. In this setup, UCNs that have the magnetic moment flipped during storage cannot be reflected by the magnetic barrier of the shutter. They will eventually reach the walls and either be reflected, captured or up-scattered. Those which are reflected will, after some collisions, fall down through the shutter aperture towards the detector.

After reaching equilibrium, the number of trapped neutrons, $N_T(t)$, and the number spin flipped neutrons that leaked and were detected, $N_L(t)$, can be described by the following set of differential equations

$$\dot{N}_T(t) = -\lambda_n N_T(t) - \lambda_{SF} N_T(t) \qquad (1)$$

$$\dot{N}_L(t) = \epsilon \lambda_{SF} N_T(t) \qquad (2)$$

where $\lambda_n$ and $\lambda_{SF}$ are respectively the neutron decay constant and the spin-flip constant and $\epsilon$ is the probability for a neutron to leave the trap after its spin has flipped. These equations assume that the time to exit the trap for spin-flipped neutrons is much smaller than the neutron lifetime. Assuming that the value for $\epsilon$ is constant, one can solve these equations for the decay constants

$$\lambda_n = \lambda_S - \lambda_{SF} = \lambda_S(1 - \alpha_{SF}) \qquad (3)$$

where $\lambda_S$ is the storage constant and $\alpha_{SF}$ is the correction associated to spin-flip. With the initial conditions, $N_T(0)=N_0$ and $N_L(0)=0$, one gets, after a storage time $T$,

$$\lambda_S = \frac{1}{T}\ln\left[\frac{N_0}{N_T(T)}\right] \qquad (4)$$

and

$$\alpha_{SF} = \frac{1}{\epsilon} \cdot \frac{N_L(T)}{N_0 - N_T(T)} \qquad (5)$$

One can see from Eq. (5) that the sensitivity of the correction $\alpha_{SF}$ to the collection efficiency of the spin-flipped neutrons, $\epsilon$, is driven by a ratio which is suppressed for $N_L(T) \ll N_0 - N_T(T)$. In other words, when the leaks are small, the precision requirements on the collection efficiency become also weaker.

Two campaigns of measurements, referred to below as runs A and B, have been carried out with the mag-netic trap described above. The duration of each campaign was a full 50 days cycle of the ILL reactor.

For run A, the aperture of the UCN guide inside the magnetic shutter was ∅20 mm. The outer solenoid was here used to eliminate zero field regions in the trap but not to induce the forced spin-flip. Due to the size of the magnetic shutter aperture, the time to remove the fast UCNs from the trap, referred to as "cleaning time", was of the order of 1200 s. The longer cycles resulted then in a loss of statistical sensitivity. A specific sequence was therefore set up to power the magnetic shutter after filling, in order to reduce the cleaning time. After moving down the lift inside the trap, the magnetic shutter was powered for 250 s with low current to produce a reduced magnetic barrier, and then powered to full current. An additional 100 s waiting interval was adopted before the first emptying such as to clean the neutrons that were accelerated by the increase of magnetic field. Once the shutter was fully powered, the magnetic trapping was so efficient that the average rate of leaking neutrons was a factor of 1.4 larger than the background rate of $2.50(5) \times 10^{-3}$ s$^{-1}$. The duration of the emptying window was 250 s. Background measurements were taken after each empting during 150 s. Following carefully checks of the cleaning conditions, the reference time was selected at $t_0$=400 s after closing the magnetic shutter. Neutrons counted at $t_S$=1300 s after closing the shutter result in a storage time $T$=900 s relative to the reference time. The storage time deduced from these measurements was

$$\tau_S = 1/\lambda_S = (874.6 \pm 1.7) \text{ s}. \quad (6)$$

This is consistent with the value obtained previously when filling the trap from the bottom [13] but a factor of about 10 more precise.

For run B the aperture of the UCN guide in the magnetic shutter was ∅60 mm and the outer solenoid was here sometimes used to induce a forced spin-flip. When the forced spin-flip was applied, it produced a small permanent artificial leak during the cycle. The setting of the forced spin-flip was chosen so that the net (after background subtraction) average rate of UCNs leaking between 600 s to 2200 s storage times was $1.01(2) \times 10^{-2}$ s$^{-1}$. In run B, the magnetic shutter was fully powered after the 64th second from the beginning of the trap filling sequence [14]. The duration of the emptying interval was here 100 s.

A unique powerful feature of this experiment is the possibility to directly measure the neutron spin-flip losses in Eq. (3) from a determination of the collection efficiency of the spin-flipped neutrons. This has been described in detail in Ref. [14]. If $r_0(t)$ [resp. $r_1(t)$] designates the rate of neutrons detected when the forced spin-flip was OFF (resp. ON), then the collection efficiency can be expressed as a ratio, between rate differences, taking into account the neutron decay. For a cycle with an emptying time $t_s$, a reference time $t_0$ and a duration $\Delta T$ for the emptying interval, we have [14]

$$\epsilon = \frac{N_{SF1}(t_S)}{N_{SF2}(t_S)} \quad (7)$$

where

$$N_{SF1}(t_S) = \sum_{t_i=t_0}^{t_s} [r_1(t_i) - r_0(t_i)] e^{\lambda_n(t_i - t_0)} \quad (8)$$

is the number of neutrons that got their spin flipped during the storage interval and

$$N_{SF2}(t_S) = \sum_{t_i=t_s+1}^{t_s+\Delta T} [r_0(t_i) - r_1(t_i)] e^{\lambda_n(t_i - t_0)} \quad (9)$$

is the number of spin-flipped neutrons which are missing when the trap is emptied. Detailed checks of the cleaning conditions have been performed here as well. The application of the procedure above for a time $t_s$=1800 s after closing the shutter, with $\Delta T$=100 s and $t_0$=300 s resulted in $\epsilon$=0.90(2) [14]. A more precise value was obtained from measurements with $t_s$=1000, 1300 and 1800 s, $\epsilon$=0.903(7). However, such a precision is not required when the level of leaks in Eq. (5) is small, as was the case during run A.

As mentioned above, the cleaning time for trapped neutrons was significantly different for runs A and B. The situation is however diametrically opposite for spin-flipped neutrons because these are very quickly drained out from the trap. With the value of $\epsilon$ measured from run B it is then possible to calculate the spin-flip correction $\alpha_{SF}$, Eq. (5), to obtain the neutron lifetime. The result extracted using data from runs A and B is

$$\tau_n = (878.3 \pm 1.9) \text{ s} \quad (10)$$

where the error is purely statistical.

Equation (1) does not take into account UCN losses due to up-scattering with the residual gas. In order to estimate the effect of the residual gas, the dependence of the number of stored neutrons with the pressure, $p$, inside the trap volume was measured (Fig. 2). In the presence of losses due to the residual gas, the number of trapped UCN as a function of time becomes

$$N_T(t) = N_0 \, e^{-(\lambda_S + \lambda_p p)t} \quad (11)$$

For a storage time of 2200 s, it was obtained that $\lambda_p = 0.15(4)$ (s·torr)$^{-1}$. This means that, a relative variation of $10^{-3}$ on $\lambda_n \approx \lambda_S \approx 1.1 \times 10^{-3}$ s$^{-1}$ corresponds to a pressure level of $10^{-5}$ torr. During the experiments, the pressure in the volume of the magnetic trap was of the order of $1.1 \times 10^{-6}$ torr.

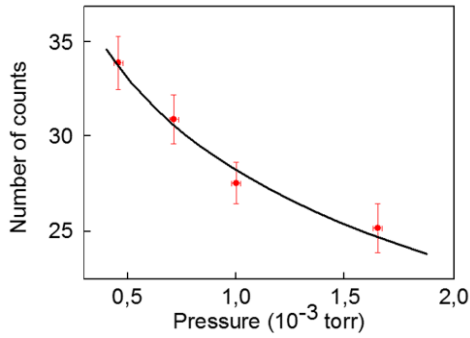

FIG. 2. Variation of the number of UCN after a storage time of 2200 s as a function of the pressure of the residual gas.

As already mentioned, the surfaces of the trap magnets were covered with Fomblin grease. In a separate measurement, the partial pressure of Fomblin vapor was investigated with a quadruple mass spectrometer [15]. It was shown that, at the pressure used during the measurements, the partial pressure of Fomblin vapor is at the level of $2 \times 10^{-9}$ torr and its effect can be neglected at the current level of precision.

No systematic effect has been found at the present level of precision. The result quoted in Eq.(10) is the most precise measurement of the neutron lifetime using magnetically stored ultra-cold neutrons. Figure 3 shows the values of the neutron lifetime included by the PDG in the current average [4]. The filled squares are from experiments with neutron beams, the filled circles were obtained with traps, and the open circle is the value from this work. The present result is consistent with the current PDG average, which includes a scale factor of 1.9, and with previous results obtained using stored UCNs but is at variance with the result obtained using a neutron beam [6].

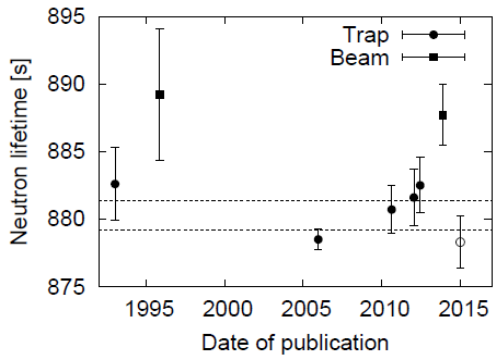

FIG. 3. Comparison of the value for the neutron lifetime obtained from this work (open circle) with the values included in the current PDG average (filled squares and circles). The dotted lines indicate the ±1σ limits of the current average. See Ref. [4] for details.

A high precision measurement using the technique of magnetically trapped UCNs is feasible. A new magnetic trap made of permanent magnets, with a 9 times larger volume, is presently under construction by our collaboration in order to reach at precision level of 0.3 s.

We are indebted to T. Brenner for his assistance during the experiment. We thank F.J. Hartman, A. Mueller, S. Paul, R. Picker, A. Serebrov and O. Zimmer for their contributions at an earlier stage of this project and E. Kats for useful discussions. This work was supported in part by the Russian Foundation for Basic Research (grant Nr 13-02-00402-a), by the French Ministry of Foreign Affairs (ECO-NET action 12649SH) and by the French Agence Nationale de le Recherche (Projet BLANC-06-3-136829 "Onelix").

———————

(*) Present address: National Superconducting Cyclotron Laboratory, Michigan State University, East Lansing MI, USA.